\documentclass[apjl]{emulateapj}
\usepackage{apjfonts}
\journalinfo{To appear in ApJ, 607, L123 (June 1, 2004)}
\submitted{To appear in ApJ, 607, L123 (June, 1, 2004)}
\shorttitle{DYNAMICAL FRICTION NEAR THE GALACTIC CENTER}
\shortauthors{KIM, FIGER, \& MORRIS}
\def\spose#1{\hbox to 0pt{#1\hss}}
\newcommand\lsim{\mathrel{\spose{\lower 3.0pt\hbox{$\mathchar"218$}}
     \raise 2.0pt\hbox{$\mathchar"13C$}}}
\newcommand\gsim{\mathrel{\spose{\lower 3.0pt\hbox{$\mathchar"218$}}
     \raise 2.0pt\hbox{$\mathchar"13E$}}}
\newcommand\msun{{\rm \,M_\odot}}

\newcommand\HeI{He~{\small I} }
\begin{document}
\title{DYNAMICAL FRICTION ON GALACTIC CENTER STAR CLUSTERS \\
WITH AN INTERMEDIATE-MASS BLACK HOLE}
\author{Sungsoo S. Kim,\altaffilmark{1} Donald F. Figer,\altaffilmark{2}
and Mark Morris\altaffilmark{3}}
\altaffiltext{1}{Dept. of Astronomy \& Space Science, Kyung Hee University,
Yongin-shi, Kyungki-do 449-701, Korea; Also at Institute of Natural Sciences,
Kyung Hee University; sskim@ap.khu.ac.kr}
\altaffiltext{2}{Space Telescope Science Institute, 3700 San Martin Drive,
Baltimore, MD 21218; figer@stsci.edu}
\altaffiltext{3}{Dept. of Physics \& Astronomy, University of
California, Los Angeles, CA 90095-1562; morris@astro.ucla.edu}

%%%%%%%%%%%%%%%%%%%%%%%%%%%%%%%%%%%%%%%%%%%%%%%%%%%%%%%%%%%%%%%%%%%%%%%%%%%%%%%%
\begin{abstract}
Numerical simulations of the dynamical friction suffered by a Galactic
center star cluster harboring an intermediate-mass black hole (IMBH) have
been performed.  Gerhard has suggested that dynamical friction, which
causes a cluster to lose orbital energy and spiral in toward the Galactic
center, may explain the presence of a cluster of very young stars in the
central parsec, where star formation might be prohibitively difficult
because of strong tidal forces.  However, numerical simulations by
Kim \& Morris showed that this is only possible if the cluster initially
has an extremely dense core.  Hansen \& Milosavljevi\'c recently suggested
that the presence of an IMBH in the cluster core might stabilize the
core against tidal disruption during the inspiral through dynamical
friction, and thus might easily deliver young stars down to the central
parsec.  We find that the presence of an IMBH does lower the minimum
initial core density required to transport young stars down to the
central parsec, but this is possible only when the mass of the IMBH
is at least $\sim 10$~\% of the total cluster mass.  This fraction
is significantly higher than that estimated by Portegies Zwart \&
McMillan with numerical simulations of IMBH formation by successive
merging of stars in the cluster core, so it does not appear that a
realistic IMBH can help transport young stars into the central parsec.

\end{abstract}
\keywords{stellar dynamics --- Galaxy: center ---  Galaxy: kinematics and
dynamics --- galaxies: star clusters --- methods: N-body simulations}

%%%%%%%%%%%%%%%%%%%%%%%%%%%%%%%%%%%%%%%%%%%%%%%%%%%%%%%%%%%%%%%%%%%%%%%%%%%%%%%%
\section{INTRODUCTION}
\label{sec:introduction}

The central parsec of the Galaxy contains a cluster of very young stars,
including $\sim 16$ very luminous \HeI emission line stars
(Krabbe et al. 1995; Paumard et al. 2001) as well as many O and B
stars (Eckart, Ott, \& Genzel 1999).  Krabbe et al. (1995) find that
the properties of these stars can be accounted for by a burst of star
formation between 3 and 7~Myr ago.  The \HeI stars appear to be evolved
massive ($> 40 \msun$) stars with stellar ages of $\sim 5$~Myr
(Paumard et al. 2001).
Krabbe et al. (1995) estimate that the number of OB stars with
$L \ge 3 \times 10^5 \, {\rm L_\odot}$ and Wolf-Rayet stars
(WNL, WCL, \& \HeI stars) in the central parsec is $\sim 50$.
We define these objects as Young Massive Stars (YMSs), and estimate
their progenitor masses to be $\gsim 40 \msun$.
Despite their very young ages, {\it in situ}
formation of these stars may be inhibited by the strong
tidal forces in the central parsec.  It is not clear that the maximum 
density of gas currently in the central parsec can be as high as the 
Roche density required for a cloud to remain bound, $\sim 4~\times~10^8~  
{\rm H \, cm^{-3}} (1~{\rm pc}/R)^3$ for galactocentric radius $R$.
Jackson et al. (1993) report 
a density of a few times $10^6 \, {\rm cm^{-3}}$ at a distance of 
$\sim 1$~pc, while Christopher et al. (2004) argue that densities 
approaching the limiting Roche density can be found at a
distance of $\sim 2$~pc; (see the discussion in \S~\ref{sec:results} below)

One possibility is that the gravitational collapse leading to the
formation of the present cluster of young stars in the central parsec
was triggered by infall of a particularly dense gas cloud, which
experienced compression by shocks involving cloud-cloud collisions,
self-intersecting gas streams, or violent explosions near or at the
central black hole (Morris 1993; Sanders 1998; Morris, Ghez \& Becklin 1999).
Alternatively, the star cluster could have formed outside the central
parsec, where tidal forces are relatively weaker and star formation is
consequently less problematic, and later migrated into the Galactic
center (GC).  Gerhard (2001) proposes that dynamical friction can bring
a massive young star cluster, initially embedded in its parent molecular
cloud, into the central parsec during the lifetime of its most massive
stars, depending on the initial location and mass of the cluster.
The drag force represented by dynamical friction, acting in the
direction opposite to the cluster motion, is owed to the induced
``wake'' of background stars.  If the star cluster is massive enough,
the resulting deceleration can in principle be large enough to cause
the cluster to spiral into the GC.

Kim \& Morris (2003; Paper I hereafter) performed numerical simulations
of dynamical friction suffered by star clusters near the Galactic center,
and found that dynamical friction can indeed bring star clusters formed
outside the central parsec into the central parsec.  However, this is
only possible if the cluster is either very massive ($\sim 10^6 \msun$) or
is formed near the central parsec ($\lsim 5$~pc). In both cases, the cluster
should have an initally very dense core ($\sim 10^8 \msun {\rm pc}^{-3}$).
These extreme requirements make the dynamical friction scenario rather
implausible.  Clusters with smaller masses, larger galactocentric
radii, and/or smaller core densities either 1) completely evaporate before
reaching the central parsec due to increasingly strong tidal forces that the
cluster must endure during the inward migration, or 2) do not reach the
central parsec within the lifetime of young, massive stars.

McMillan \& Portegies Zwart (2003) presented semianalytic calculations
of the inspiral of star clusters near the GC to study the parameter space
in which the cluster can reach the central parsec of the Galaxy within
a few million years.  While they performed simplified semianalytic
calculations to explore a wide range of parameter space, Paper I
implemented a numerical treatment that models both the cluster and the inner
part of the Galaxy with a large number of particles to accurately model
the final dissolution phase of the cluster.  The amount of mass
deposited in the central parsec can be estimated more accurately in this
way, and this estimation was the main goal of Paper I.

Recently, Hansen \& Milosavljevi\'c (2003) suggested that if a cluster
formed outside the central parsec harbors an intermediate-mass
($10^3$--$10^4 \msun$) black hole (IMBH),\footnote{We do not address
in the present study how such an IMBH would be formed.}
the dynamical friction scenario
may work with more plausible initial cluster conditions.  The idea is
that the deep potential well induced by the IMBH at the center of the
cluster may be able to keep the cluster core intact against tidal disruption
for a longer time during the inward migration by dynamical friction
so that the cluster core can reach the central parsec.  However, it is
difficult to determine with the (semi)analytic approach of Hansen \& 
Milosavljevi\'c (2003) 1) how long the IMBH can hold the cluster core 
intact, and 2) how much stellar mass would ultimately be deposited into
the central parsec after cluster dissolution.  The present study
carries out numerical simulations of dynamical friction on GC
star clusters having an IMBH using the same numerical method as in Paper I
in order to answer these questions.

%Our models and the method of simulation are described in 
%\S~\ref{sec:models}, and the simulation results are presented and
%discussed in \S~\ref{sec:results}.  Our findings are then summarized
%and discussed in \S~\ref{sec:discussion}.

%%%%%%%%%%%%%%%%%%%%%%%%%%%%%%%%%%%%%%%%%%%%%%%%%%%%%%%%%%%%%%%%%%%%%%%%%%%%%%%%
\section{MODELS}
\label{sec:models}

We use the same numerical models as in Paper I, except that the
clusters in the present study harbor an IMBH at the center.
Here we briefly describe our models, and readers are referred
to Paper I for a detailed model description.

We use a parallelized N-body/SPH (Smoothed Particle Hydrodynamics) code
named {\sc Gadget} (Springel, Yoshida \& White 2001), which computes
gravitational forces with a hierarchical tree algorithm and represents
fluids by means of SPH.  We implement only the gravitational part of the code.

To model the potential of the central region of the Galaxy, we adopt
a truncated, softened, spherical, power-law density profile:
\begin{equation}
	\rho_g = {4 \times 10^6 \over 1+(R/0.17\,{\rm pc})^{1.8}} \,
		 \exp(-(R/R_{trunc})^6) \, \msun {\rm pc^{-3}}.
\end{equation}
This is a density model from Genzel et al. (1996) with an added exponential
truncation, scaled by $R_{trunc}=15$~pc.  The outer boundary of the Galaxy
is set to be 25~pc.  The net angular momentum of the Galaxy is assumed to
be zero.  We use $8.4 \times 10^5$ particles to model the Galaxy, and
the total Galaxy mass represented in our model is $4.3 \times 10^7
\msun$.\footnote{$M_{galaxy}$ for Models 2 \& 3 in Table 1 of Paper I are
erroneous and should read $4.3 \times 10^7 \msun$ and $6.0 \times 10^7 \msun$,
respectively.}
Thus the mass represented by a single Galaxy particle is $\sim 50 \msun$.

For the initial stellar density profile of the cluster, we adopt the Plummer
density profile,
\begin{equation}
	\rho_{cl} = {3 M_{cl} \over 4 \pi r_c^3} \left ( 1 +
		   {r^2 \over r_c^2} \right )^{-5/2}.
\end{equation}
Clusters in our simulations initially have a total cluster mass $M_{cl}$
of $10^5 \msun$, galactocentric radius $R$ of 5~pc, core radius $r_c$ of
0.13~pc, and tidal radius $r_t$ of 0.77~pc (thus $r_t/r_c=6$).
The number of particles for the cluster is $10^4$,
making the mass of each cluster particle $\sim 10 \msun$ (this number
is not exactly $10 \msun$ because $M_{cl}$ includes the mass of the IMBH,
which is represented by a single particle).

%Table 1
\begin{deluxetable}{ccccc}
%\scriptsize
\tablecolumns{10}
\tablewidth{0pt}
\tablecaption{Simulation Parameters
\label{table:sim}}
\tablehead{
\colhead{} &
\colhead{$v_{init}$} &
\colhead{$M_{IMBH}$} &
\colhead{$\rho_c$} &
\colhead{$r_{in}$} \\
\colhead{Simulation} &
\colhead{($v_{circ}$)} &
\colhead{($\msun$)} &
\colhead{($\rm \msun pc^{-3}$)} &
\colhead{(pc)}
}
\startdata
1 &  0.5 &  $            0$ &  $11.4 \times 10^7$ &  0.00  \nl
2 &  0.5 &  $2 \times 10^4$ &  $5.5  \times 10^6$ &  0.06  \nl
3 &  0.5 &  $1 \times 10^4$ &  $7.2  \times 10^6$ &  0.05  \nl
4 &  0.5 &  $3 \times 10^3$ &  $8.7  \times 10^6$ &  0.04  \nl
5 &  0.2 &  $1 \times 10^4$ &  $7.2  \times 10^6$ &  0.05  \nl
\enddata
\end{deluxetable}

The Plummer density profile we adopted above is only for the
stellar component, and we place an IMBH at the center of this
profile.  However, an equilibrium configuration may not always be achievable
when an IMBH is added to this profile (a part of the distribution function
$f(E)$ may become negative if the IMBH mass added is too large; see Binney \&
Tremaine 1987 for details).  It turns out that for the mass
range of the IMBHs we consider in the present study, equilibrium is not
available without modifying the adopted stellar density profiles.
We find that equilibrium is achievable when stars inside a certain
small radius from the cluster center, $r_{in}$, are removed
(once the simulation begins, the initial void of stars near the IMBH will
be quickly filled by stars that are deflected into the void by two-body
interactions, and the cluster will try to reach the new equilibrium).
The values of $r_{in}$ that we choose are the minimum radii that allow
equilibrium
for the system, and are shown in Table~\ref{table:sim}.  Each simulation
has a different initial central density ($\rho_c$) because of different
$r_{in}$ values.

The initial orbital eccentricity is expressed in terms of the initial
orbital velocity, $v_{init}$, relative to the circular velocity at
a given $R$, $v_{circ}$, and our clusters initially have a tangential
velocity only.
Our simulations have $v_{init}/v_{circ}$ of either 0.5 or 0.2.  Such
significantly non-circular initial orbits might result from
cloud-cloud collisions, which is one of the possible mechanisms for
forming a relatively massive cluster near the central parsec.  Such highly
eccentric initial orbits will make the dynamical friction more effective,
as the cluster will experience its largest tidal forces near its periapse 
position.  Since this assumption favors the process we are investigating,
it strengthens our conclusion in what follows that the process is unlikely to  
be operating in the GC.  

As discussed in Paper I, clusters with $\gsim 10^6 \msun$ would
initially contain more than a thousand YMSs,
which is more than an order of magnitude larger than currently observed
in the central parsec (there are very few YMSs outside this
region except for two very young star clusters at least 30~pc
from the GC with a mass of order of $10^4 \msun$,
the Arches cluster and the Quintuplet
cluster\footnote{See Paper I for references to work on these clusters}).
%Thus observing only $\sim 16$ \HeI stars after the birth of
%$\gsim 10^6 \msun$ cluster would be possible only during a very narrow
%window of time.  Furthermore, estimated masses of the Arches and
%the Quintuplet clusters (Kim et al. 2000), and the molecular clumps
%in the CND (Christopher et al. 2004) are all $\lsim 10^5 \msun$.
%Therefore we consider $10^5 \msun$ clusters in the present study.
On the other hand, if $M_{cl}$ is less than $10^4 \msun$, the number of
YMSs in the cluster would be only a fraction of what
is currently observed in the central parsec, and one would need
successive inspirals of several clusters to explain the observed number
of YMSs.  Thus we choose $M_{cl}$ = $10^5 \msun$ for our
clusters.

%%%%%%%%%%%%%%%%%%%%%%%%%%%%%%%%%%%%%%%%%%%%%%%%%%%%%%%%%%%%%%%%%%%%%%%%%%%%%%%%
\section{RESULTS}
\label{sec:results}

\begin{figure}
%Fig 1
\centerline{\epsfxsize=8.0cm\epsfbox{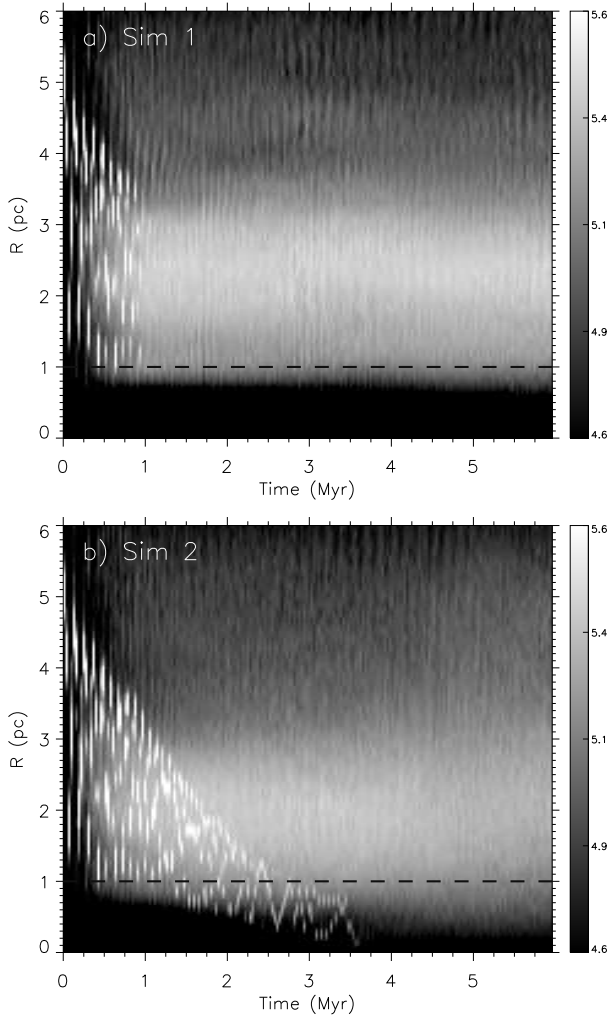}}
\caption
{\label{fig:compare}Grey scale map showing the temporal evolution of the 
radial distribution of cluster particles for Simulations
1 and 2.  The IMBH is not present in Simulation~1 but is included in
Simulation~2, though it is not shown in the plot.  The grey scale bars
next to the map represent the logarithmic scales of the density in units
of $\msun {\rm pc^{-1}}$.  The horizontal dashed line shows the location
of $R = 1$~pc.  In Simulation~2, the core (relatively bright, distinct
spots) continues to sprial in below $R = 1$~pc.
}
\end{figure}

Figure~\ref{fig:compare} compares the evolution of the radial
distribution of cluster particles of Simulation~1 (Simulation 8 of Paper I)
and Simulation~2, which have the same initial conditions except that
the former does not have an IMBH.  The orbital evolution shows a periodic 
oscillation simply because the clusters initially have elliptical orbits.
This plot dramatically shows the role of the IMBH in making the cluster core
more stable against tidal disruption and prolonging its inward
migration to smaller galactocentric radius.  While the whole
cluster of Simulation~1 completely disrupts when reaching $R=1$~pc and
is spread over the region between 0.8 and 3.5~pc, in the case of Simulation~2, 
where the cluster harbors an IMBH at its center, only the halo of the cluster
disrupts outside the central parsec and the core continues to spiral in
to the central parsec.  Although the location of the IMBH is not plotted
in this Figure, we find that the IMBH and the core are tightly bound to
each other until the core is dissolved inside the central parsec at
$\sim 3.5$~Myr.

Paper I showed that without the IMBH, a $10^5 \msun$ cluster must have
$\rho_c$ of at least $\sim 10^8 \msun {\rm pc}^{-3}$ to deliver its core
to the central parsec within the lifetime of the YMSs, but Simulation~2
lowers this requirement by almost two orders of magnitude.  The initial
$\rho_c$ of Simulation 2, $5.5 \times 10^6 \msun {\rm pc}^{-3}$, can often 
be observed in the cores of dense globular clusters, and is close to
the density estimated for the early phase of the Arches cluster
(Kim et al. 2000).

\begin{figure}
%Fig 2
\centerline{\epsfxsize=8.0cm\epsfbox{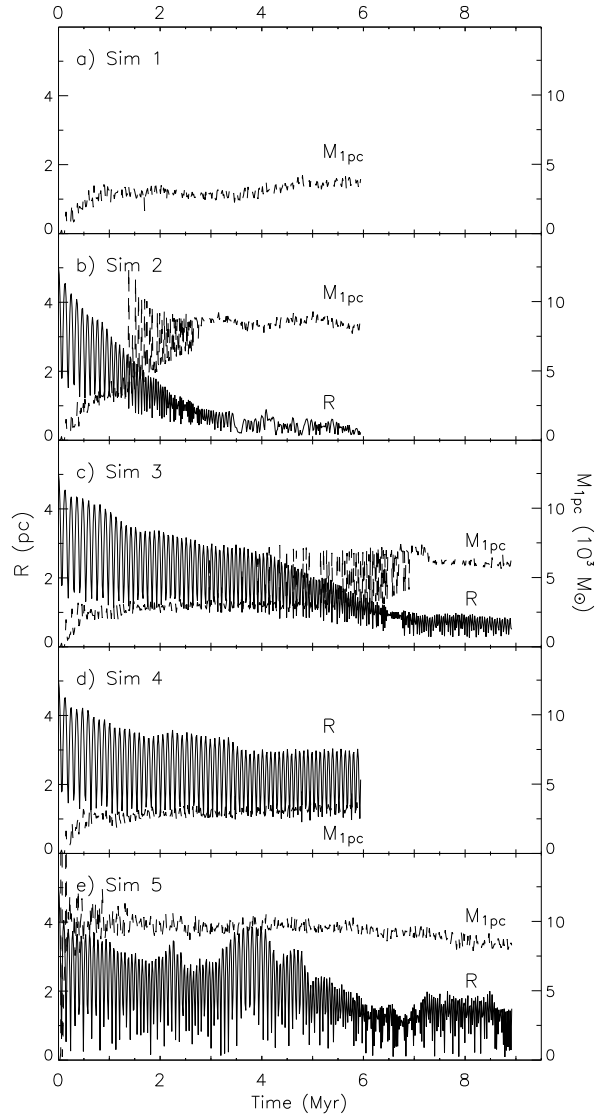}}
\caption
{\label{fig:main}Evolution of galacocentric radius, $R$, of the IMBH
({\it solid line}) and the mass of the cluster stars that are located
in the central parsec of the Galaxy, $M_{\rm 1pc}$ ({\it dashed line}).
}
\end{figure}

Figure~\ref{fig:main}b shows the location of the IMBH and the mass
of the cluster stars (not including the IMBH) located inside
the central parsec, $M_{1pc}$, as a function of time for Simulation~2.
$M_{\rm 1pc}$ increases as the IMBH enters the central parsec, and reaches
its asymptotic value when the cluster core is completely dissolved.
The stellar mass deposited in the central parsec is $\sim 8 \times 10^3 \msun$
for Simulation~2, which is approximately three times larger than that
of Simulation~1 (Figure~\ref{fig:main}a).
Clusters formed near the GC are expected to undergo very
rapid dynamical mass segregation during the first few $10^5$~yr
(Kim, Morris, \& Lee  1999).  Thus by the time the core sinks to the
central parsec and dissolves therein, the core will be largely dominated
by the most massive stars.  If we assume that the typical mass of a YMS
is $50 \msun$ and the number of central parsec stars presumed to be YMSs
is 50, then the total mass of YMSs in the central parsec is $2.5 \times
10^3 \msun$.  This value is sufficiently less than $M_{\rm 1pc}$ at the end of
the simulation that, allowing for a reasonable initial mass function,
we expect that almost all YMSs formed in the cluster
of Simulation~2 are transported into the central parsec.

$M_{\rm 1pc}$ at the end of Simulation~1 is $\sim 2.5 \times 10^3 \msun$,
which equals the estimated mass of YMSs in the central parsec.  But only a
few stars of Simulation~1 are located inside $R=0.8$~pc, while most of
the observed YMSs are located inside $R=0.8$~pc.  The presence of
the IMBH in Simulation~2 not only increases $M_{\rm 1pc}$, but also
transports the core stars deeper into the GC, as shown in 
Figure~\ref{fig:compare}.

It may appear that Simulation~2 is a good scenario to explain the
presence of YMSs in the central parsec, but there is
nonetheless a rather serious problem with it: 
the IMBH in Simulation~2 is too massive compared to the total cluster 
mass (20~\% of the cluster mass).  The typical core
mass in a cluster is $\sim 1$~\% of the cluster mass, so even if all
the stars in the cluster core turn into an IMBH, it is still far smaller
than the IMBH of Simulation~2.  Furthermore, Portegies Zwart \& McMillan
(2002) estimate that an object resulting from the runaway growth through
stellar collisions can grow only up to $\sim 0.1$~\% of the cluster mass.
The IMBH could grow further by accreting remnant gaseous material left 
over from the cluster formation before the gas gets blown away by massive 
stars, but it seems highly unlikely that there would be enough gaseous 
material for the IMBH mass to grow by a factor
of 10 to 100 by accretion in just a few million years.

Clusters having IMBHs smaller than that of Simulation~2 would be more
realistic, but then the stabilization of the cluster core by the IMBH
would be less effective.  By comparing Simulations 2, 3, and 4 
(Fig.~\ref{fig:main}b, c, \& d), which have the same initial conditions
except for the IMBH mass, we find that the minimum IMBH mass fraction
required to bring the core of the $10^5 \msun$ cluster to the central
parsec is $\sim 10$~\% ($\sim 10^4 \msun$).  $M_{\rm 1pc}$ at the
end of Simulation~4 is almost the same as that of Simulation~1,
where there is no IMBH.  It takes more than 6~Myr for the core of
Simulation~3 to migrate to the central parsec and dump the stars there,
and this time is close to the upper limit of the estimated age of the
YMSs in the central parsec.

%Figure~\ref{fig:main}e shows that the core of a cluster with an initial
%$\rho_c$ of $4.8 \times 10^5 \msun {\rm pc}^{-3}$, which is 8 times
%smaller than that of Simulation~2, disrupts before reaching the central
%parsec.  This implies that for a cluster with $M_{cl}=10^5 \msun$
%and $M_{IMBH}=2 \times 10^4 \msun$, the minimum $\rho_c$ required to
%deliver the core to the central parsec is $\sim 10^6 \msun {\rm pc^{-6}}$.

A more eccentric initial cluster orbit could be more effective for
the inward migration of the cluster core because the cluster would
experience higher Galactic stellar densities at earlier times.
However, Simulation~5 (Figure~\ref{fig:main}e), which has
$v_{init}/v_{circ}=5$, shows that $M_{\rm 1pc}$ is not sensitively
dependent on the initial eccentricity of the cluster orbit.

We also performed a couple of simulations with an initial density profile
from the King model (King 1966) with a concentration parameter $W_0=9$,
which represents a highly concentrated profile for a cluster, but found
that the results are nearly identical to the simulations with the Plummer
initial profiles.  This shows that our results are probably not sensitive
to the choice of the initial density profile.
Initial $R$ values other than 5~pc were not tried in the present study,
but as found in Paper I, survival of the cluster depends on the initial
$\rho_c$, but not sensitively on the initial $R$.

After trying and considering various different initial conditions for
the cluster, we conclude that the presence of an IMBH that occupies less
than $\sim 10$~\% of the total cluster does not greatly increase the
amount of stars deposited in the central parsec.

The simulation method adopted in the present study properly describes
dynamical friction, but may not accurately follow internal dynamical
evolution of star clusters due to the use of a rather large softening
parameter (see Paper I), as well as to the uniformity of the model stellar
masses, which cannot therefore reproduce mass segregation in the cluster.
The two-body relaxation times at the
core of our clusters are of order of $10^5$~yr, so some cluster cores may
experience significant dynamical evolution during our simulation periods.
In star clusters with a black hole of a considerable mass, the core density
gradually decreases by the loss of stars to the black hole and the expansion
of the core (Rauch 1999, Freitag \& Benz 2002).  This density decrease
is probably underestimated in our simulations, thus the actual initial core
densities and black hole mass fraction required to explain the YMSs in
the central parsec are probably even higher than implied by our results.

We note that from high spatial resolution millimeter line observations of
the inner four parsecs of the GC, Christopher et al. (2004) found
that some of the molecular clumps in the circumnuclear disk ($1.2 \gsim
R/{\rm 1~pc} \gsim 2.5$) appear to be dense enough to surpass the Roche
limit in the central parsec.  The virial densities of their clumps
range between $10^7$ and $10^9 {\rm H \, cm^{-3}}$, and the virial masses
range between $10^3$ and $10^5 \msun$.
Simulations in Paper I suggest that a cluster formed from the heaviest
clump in the Christopher et al. observations (an inferred $\sim 10^5 \msun$) 
may be able to spiral into the central parsec within the lifetime of the YMSs, 
even without an IMBH.

\acknowledgements
S.S.K. thanks Holger Baumgardt, Micol Christopher, Hyung Mok Lee, Stephen
McMillan, Simon Portegies Zwart, and Nick Scoville for helpful discussion.
This work was supported by the Astrophysical Research Center for
the Structure and Evolution of the Cosmos (ARCSEC) of Korea Science
and Engineering Foundation through the Science Research Center (SRC) program.
Simulations presented in this {\it Letter} were performed on the linux
cluster at Korea Astronomy Observatory (KAO), which was built with the
fund from KAO and ARCSEC.  We are grateful to Jongsoo Kim for his kind
assistance with the linux cluster.

%%%%%%%%%%%%%%%%%%%%%%%%%%%%%%%%%%%%%%%%%%%%%%%%%%%%%%%%%%%%%%%%%%%%%%%%%%%%%%%%

\end{document}